\documentclass[aps,prb,twocolumn,nobibnotes]{revtex4-2}  
\usepackage{dcolumn}
\usepackage{graphicx}
\usepackage{color}
\usepackage{amssymb}
\usepackage{amsmath}
\usepackage{bm}
\usepackage[colorlinks=true,linktocpage=true,breaklinks=true]{hyperref}
\usepackage{multirow}
\usepackage{braket}
\usepackage{appendix}

\begin{document}
\title{Impact of Magnon Interactions on Transport in Honeycomb Antiferromagnets }
\author{Konstantinos Sourounis}
\author{Aurélien Manchon}
\email{aurelien.manchon@univ-amu.fr}
\affiliation{Aix-Marseille Université, CNRS, CINaM, Marseille, France}
\begin{abstract}
The thermal transport of magnons has attracted substantial attention as an energy-efficient alternative to the transport of electrons. Most theoretical studies so far have been carried out within the frame of the linear spin-wave theory, which dramatically fails upon increasing the temperature and in the presence of competing interactions. In this work, we consider the impact of three- and four-magnon interactions in a honeycomb antiferromagnet, where such interactions are remarkably strong even at zero temperature. Using a combination of quantum field theory and mean-field theory, we compute the band structure of the interacting magnons and investigate the spin Nernst effect. We find that in the presence of in-plane Dzyaloshinskii-Moriya Interaction, the three-magnon interaction induces a non-reciprocal band splitting, even at zero temperature, leading to an enhancement of the spin Nernst conductivity. In contrast, the four-magnon interaction renormalizes the magnon spectrum at high temperatures, leading to a reduction of the overall magnon spin Nernst effect. These results suggest that interactions can massively influence the transport properties of magnons in antiferromagnets, even at zero temperature, and should be taken into account for predictive modeling.
\end{abstract}

\maketitle
\section{Introduction}
In the field of spintronics, spin waves, and their quanta, magnons, have been considered as a viable alternative to electrons for transporting information. They present several advantages compared to electrons, including their wide range of frequencies, the existence of non-reciprocal effects, as well as, their intrinsically low energy dissipation when considering magnetic insulators \cite{kruglyak2010magnonics, chumak2015magnon,barman20212021}. Magnon transport phenomena underscore longitudinal transport in the form of the spin Seebeck effect \cite{xiao2010theory,uchida2010observation,uchida2010spin,rezende2016theory,wu2016antiferromagnetic}, the thermal Hall effect \cite{katsura2010theory,onose2010observation,matsumoto2011rotational,matsumoto2011theoretical,matsumoto2014thermal}, and the spin Nernst effect \cite{cheng2016spin,zyuzin2016magnon,zhang2022perspective}, among other proposals \cite{li2020magnonic,wimmer2020observation,shen2020magnon,guckelhorn2023observation}.
In addition, the magnonic counterparts of electronic topological materials, known as topological magnonic materials, have been proposed in a wide variety of flavors, such as Chern Insulators \cite{shindou2013topological,zhang2013topological,mook2014edge,owerre2016first}, $\mathbb{Z}_2$ topological insulators \cite{nakata2017magnonic,kondo2019three}, nodal lines \cite{li2017dirac,mook2017magnon}, Weyl \cite{mook2016tunable,li2016weyl} and Dirac magnon \cite{pershoguba2018dirac, fransson2016magnon,yuan2020dirac}, among others. Extensive reviews on this topic can be found in the recent bibliography \cite{mcclarty2022topological,bonbien2021topological,zhuo2023topological}. Despite these numerous predictions, the experimental detection of the magnon topology remains elusive since the two most instrumental methods for detecting magnons, bulk-sensitive neutron scattering \cite{chen2018topological,chen2021magnetic,nikitin2022thermal,yuan2020dirac,scheie2022dirac,sala2021van} and surface-sensitive spin pumping  \cite{cornelissen2015long,lebrun2018tunable,vaidya2020subterahertz,li2020spin}, cannot unambiguously probe the Berry curvature information. This is particularly true in the case of anomalous thermal magnon transport, where the full magnonic band structure contributes to the transport so that topological magnon states are often blurred by trivial states (see discussion in \cite{guemard2022unified}). Furthermore, at finite temperature magnon-phonon as well as magnon-magnon interactions can dramatically modify the band structure, leading to topological phase transitions upon increasing the temperature, as recently proposed in honeycomb ferromagnets \cite{PhysRevLett.117.217205,PhysRevB.95.144420,zhang2019thermal,zhang2021anomalous,pershoguba2018dirac,lu2021topological}.

Whereas most theoretical and experimental investigations have focused on ferromagnetic magnons, antiferromagnetic materials present a remarkably rich platform for the investigation of magnonic transport. As a matter of fact, antiferromagnets are currently under intense scrutiny for spintronic applications due to their notable properties, such as the absence of stray fields, high frequencies of operation, and a wide material variety \cite{jungwirth2016antiferromagnetic,baltz2018antiferromagnetic,jungwirth2018multiple,bonbien2021topological}. Anomalous magnon transport has been investigated theoretically in both collinear \cite{zyuzin2016magnon,cheng2016spin} and noncollinear antiferromagnets \cite{mook2019thermal,li2020intrinsic,mook2019spin,Goli2021}, and anisotropic magnon dispersion has been proposed recently in spin-split antiferromagnets \cite{vsmejkal2022chiral,brekke2023two,cui2023efficient}.

\begin{figure}[b]
\includegraphics[width=8.6cm,height=8.6cm,keepaspectratio]{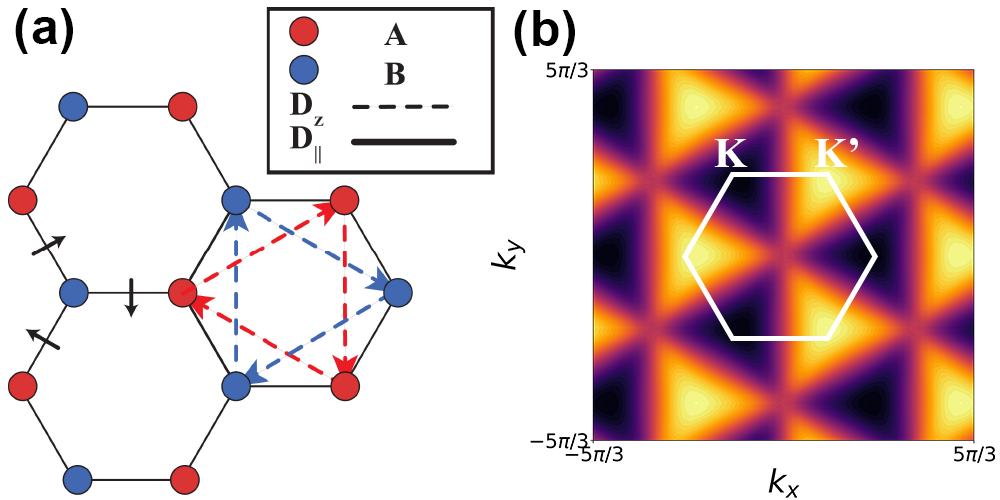}
\caption{(a) The honeycomb antiferromagnet with the Heisenberg (black) and DM (blue) interactions. (b) The Berry curvature of the non-interacting magnons}
\label{fig:fig1}
\end{figure}

Whereas theoretical research on topological thermal magnon transport has been intense over the past decade, most studies have been performed within the framework of the Linear Spin Wave Theory (LSWT), which is generally only accurate for collinear spin textures at low temperatures. Spin excitations are generally not bosonic and as soon as the temperature deviates from zero, magnon-magnon interactions emerge that massively impact the nature of the quasiparticle itself, as noted in earlier studies \cite{oguchi1960theory, harris1971dynamics}. 

At the lowest order, two classes of interactions dominate, the four-magnon interaction and the three-magnon interaction. The four-magnon interactions, associated with the magnetic exchange, have been investigated by both diagrammatic and mean field techniques and they are considered the main cause of magnetic damping both in ferromagnets and antiferromagnets \cite{zhitomirsky2013colloquium}. The three-magnon interactions are caused by the coupling of orthogonal spin components $S^zS^{\pm}$, induced by the Dzyaloshinskii-Moriya Interaction (DMI) or the non-collinear magnetic configuration itself \cite{chernyshev2009spin,zhitomirsky2013colloquium}. As such, it can both dramatically reduce the magnon lifetime and alter its Berry curvature, leading to temperature-induced topological phase transitions \cite{chernyshev2016damped,mcclarty2019non,mook2021interaction,lu2021topological,li2023temperature}. However, treating the three-magnon interactions is often mathematically complex and requires sophisticated techniques such as quantum field theory \cite{zhitomirsky2013colloquium}. More recently, a similar coupling has been proposed to capture the effect of these interactions in antiferromagnets without the troublesome calculation of self-energies \cite{matsumoto2020nonreciprocal,PhysRevLett.131.186702}. While this effect gives a good qualitative picture of the impact of in-plane DMI, it does not capture the temperature dependence and more complex physics associated with the three-magnon interactions.

In the present work, we investigate the impact of magnon-magnon interaction on the temperature dependence of the magnonic band structure and anomalous transport in a honeycomb antiferromagnet. Honeycomb antiferromagnets are known to display spin Nernst effect (SNE) \cite{cheng2016spin,zyuzin2016magnon}, topological phase transition \cite{neumann2022thermal,li2023temperature} and other interesting properties \cite{fishman2022orbital,chen2023damped}. Most importantly for the present work, magnon interactions are non-negligible even at zero temperature, which makes this system remarkable.

The article is organized as follows: In Section \ref{sectionII}, we highlight the main features of the magnon spectrum, symmetries, and magnon transport in the LSWT limit. Section \ref{sectionIII} covers the main results of this work, the impact of three- and four-magnon interactions in the honeycomb antiferromagnet. In Section \ref{sectionIIIA}, we study the four-magnon interactions that naturally rise due to the expanded Holstein-Primakoff (HP) transformation. In Section \ref{sectionIIIB}, we investigate the more complex three-magnon interactions, their spectroscopic properties, and their impact on magnon transport. In Section \ref{sectionIIIC}, we investigate the combined effects of the three- and four-magnon interactions on magnon transport. The details of the calculations, which can be sometimes rather cumbersome, are left in the Appendix. Then, in Section \ref{sectionIV}, we discuss the relevance of these effects to the experimentally available materials. Finally, in Section \ref{sectionV}, we conclude by discussing challenges in modeling interacting magnonic transport at finite temperatures.

\section{Linear Spin  Wave Theory\label{sectionII}}
\subsection{Magnon Spectrum}
We  consider the spin Hamiltonian, illustrated in  Fig. \ref{fig:fig1}(a), for the honeycomb antiferromagnet,
\begin{eqnarray}
    H &=&  J\sum_{\langle ij\rangle} {\bf S}_{A,i}\cdot {\bf S}_{B,j}+K\sum_i(S^z_i)^2 \label{eq:1}
    \\  
    &&+D_z\sum_{\langle\langle ij\rangle\rangle}{\bm z}\cdot({\bf S}_i\times {\bf S}_j)+ D_{||}\sum_{\langle ij \rangle}{\bm\eta}_{ij}\cdot( {\bf S}_{A,i}\times {\bf S}_{B,j}), \notag\end{eqnarray}
where $J$ is the nearest neighbor Heisenberg exchange, $K$ is the easy-axis anisotropy in the ${\bm z}$ direction respectively, $D_z$ is the next-nearest neighbor out-of-plane Dzyaloshinskii-Moriya interaction, A/B is the sublattice index and ${\bm \eta}_{ij}$ is the nearest neighbor vector orthogonal to the bonds [see Fig. \ref{fig:fig1}(a)]. The nearest neighbor in-plane DMI, $D_{||}$, does not enter the LSWT Hamiltonian but, instead, it introduces the three-magnon interaction that will be discussed in Section \ref{sectionIII}. Throughout this article, the magnetic anisotropy $K$ and DMI coefficients $D_z$ and $D_{||}$ are normalized to the exchange energy $J = 1$ meV. 

We use the HP transformation for antiferromagnetic magnons \cite{PhysRev.58.1098,rezende2019introduction}
\begin{eqnarray}
S^+_{A,i} &=&\sqrt{2S}a_i-\frac{a^{\dagger}_ia_ia_i}{2\sqrt{2S}}+\dots,\nonumber  \\
S^-_{A,i}&=& \sqrt{2S}a^{\dagger}_i-\frac{a^{\dagger}_ia^{\dagger}_ia_i}{2\sqrt{2S}}+\dots,\; S^z_{A,i}=S -a^{\dagger}_ia_i, \label{2.a}\\
S^+_{B,i} &=&\sqrt{2S}b^{\dagger}_i-\frac{b^{\dagger}_ib^{\dagger}_ib_i}{2\sqrt{2S}}+\dots, \nonumber \\
S^-_{B,i}&=&\sqrt{2S}b_i-\frac{b^{\dagger}_ib_ib_i}{2\sqrt{2S}}+\dots,\; S^z_{B,i}=-S+b^{\dagger}_ib_i, \label{2.b}
\end{eqnarray}
where the operators $a_i,b_i$ obey the usual boson commutation rules. Since the present section focuses on LSWT, only the lowest-order terms in the expansion are considered. Then, we apply the Fourier transform to the magnon operators $a_{\bf k},b_{\bf k}=\frac{1}{\sqrt{N}}\sum_{j}e^{i{\bf k}\cdot {\bf R}_i}a_i,b_i$ to get the Hamiltonian of our system
\begin{equation}
    H^{(2)}=  S\sum_{\bf k} \Psi_{\bf k}^{\dagger} \begin{pmatrix}
    A^+({\bf k})&B^*({\bf k})\\
    B({\bf k})&A^-({\bf k})
    \end{pmatrix} \Psi_{\bf k},\label{3}
\end{equation}
where $A^{\pm}({\bf k})=3J+K(2S-1)/S\pm D_zg({\bf k}),$ $ B({\bf k})= Jf({\bf k})$ in which $f({\bf k})= \sum_{\langle ij \rangle}\exp{(i{ \bf k}\cdot {\bm\delta^{(1)}_{ij}}})$ and $g({\bf k})=2\sum_{\langle\langle ij \rangle\rangle}\sin({\bf k}\cdot {\bm\delta}^{(2)}_{ij})$ where ${\bm\delta^{(l)}_{ij}}$ are the $l$-th nearest neighbors vectors, the magnon spinors are $\hat{\Psi}_{\bf k}=[b_{\bf k},a^{\dagger}_{\bf -k}]^T$. Here we have used the reduced basis, as described in Appendix A, and we note $K'=K(2S-1)/S$.

To diagonalize the above Hamiltonian one can use either an analytical or a numerical method \cite{colpa1978diagonalization,rezende2019introduction}. The basic equation of diagonalization is described by
\begin{equation}
    \hat{T}^{\dagger}_{\bf k}\hat{H}_{\bf k}\hat{T}_{\bf k}=\hat{\epsilon}_{\bf k}, \label{4}
\end{equation}
where $\hat{T}_{\bf k}$ is the paraunitary matrix that satisfies the identity $\hat{T}^{\dagger}_{\bf k}\hat{G}\hat{T}_{\bf k}=\hat{G}$ and the magnon spectrum is $\hat{\epsilon}_{\bf k}=(\omega_{-,{\bf k}},\omega_{+,-{\bf k}})$, as described in Appendix A. Equation \eqref{4} describes a Bogoliubov transformation of the magnon basis and, for the reduced matrix, it has the form $\hat{\Psi}_{\bf k} = \hat{T}_{\bf k}\hat{X}_{\bf k}$ or
\begin{equation}
    \begin{pmatrix}
        b_{\bf k}\\
        a^{\dagger}_{\bf -k}
    \end{pmatrix} =
       \begin{pmatrix}
        T_{b-}&T_{b+}\\
        T_{a-}^*&T_{a+}^*
    \end{pmatrix}
       \begin{pmatrix}
        c_{-,{\bf k}}\\
        c^{\dagger}_{+,-{\bf k}}
    \end{pmatrix}, \label{5}
\end{equation}
where $\hat{X}_{\bf k}=[c_{-,{\bf k}},c^{\dagger}_{+,-{\bf k}}]$ is the reduced spinor in the diagonalized basis.

\subsection{Berry curvature and symmetries}

Using the analytical gauge-fixed expression of the eigenvectors \cite{xiao2010berry}, we compute the Berry curvature \cite{cheng2016spin}
\begin{equation}
    \Omega^{xy}_{n,{\bf k}}= i\bra{\nabla_{k_x}T_{n,{\bf k}}}G\ket{\nabla_{k_y}T_{n,{\bf k}}}-(x\leftrightarrow y). \label{6}
\end{equation}

The Heisenberg and easy axis anisotropy terms in Hamiltonian \eqref{eq:1} are invariant under the $\mathcal{T}C_2$ symmetry, whereby $C_n$ is the rotation symmetry by a $2\pi/n$ angle, but break the inversion symmetry $\mathcal{I}$ due to the existence of the two sublattices, A and B, giving rise to a Berry curvature even in the absence of DMI. Both of these symmetry properties are illustrated in the Berry curvature in  Fig. \ref{fig:fig1}(b). $D_z$ breaks the $\mathcal{T}C_2$ symmetry but does not enter the paraunitary matrix $\hat{T}_{\bf k}$ due to its commutation with the $\hat{G}$ matrix and, as such, it does not directly influence the Berry curvature \cite{cheng2016spin,zyuzin2016magnon}.

The in-plane DMI, $\propto D_{||}$,  also breaks the $\mathcal{T}C_2$ symmetry but to account for its effects, one needs to consider nonlinear magnon terms. Interestingly, because its direction depends on the vectors ${\bm \eta}_{ij}$, the in-plane DMI mediated by three-magnon interaction induces a phase $e^{i\phi_{ij}}$, where $\phi_{ij} = -\frac{2n\pi}{3}$ ($n$=1,2,0) that promotes a valley-dependent band splitting, as discussed in Section \ref{sectionIIIB}. Additionally, due to the geometry of the honeycomb lattice, the in-plane DMI favors the transverse motion of magnons, resulting in enhanced spin Nernst angles as discussed further below.

\subsection{Magnon Transport}
We investigate the magnon transport induced by a thermal gradient by addressing the magnon Seebeck effect (i.e., the longitudinal magnon transport) and the magnon SNE (i.e., transverse pure spin transport), expressed as \cite{cheng2016spin,zyuzin2016magnon,rezende2019introduction,zhang2022perspective} 
\begin{eqnarray}
    \sigma_{xx} &=& \frac{1}{2\hbar \alpha_0  k_BT^2}\sum_n^N\int d{\bf k} v_{x,n}^2\frac{e^{\hbar\epsilon_{n,{\bf k}}/k_BT}}{(e^{\hbar\epsilon_{n,{\bf k}}/k_BT}-1)^2}, \label{eq:7}\\
    \sigma_{xy} &=&\frac{k_BT}{\hbar}\sum_n^N\int d{\bf k} \Omega_{n,{\bf k}}^{xy}\left( c_1(n_b(\epsilon_{n,{\bf k}}))-c_1(n_b(\epsilon_{n,-{\bf k}}))\right),\nonumber\\\label{eq:8}
\end{eqnarray}
where $\alpha_0$ is the phenomenological damping constant, the magnon velocity is $v_{x,n}=\partial \epsilon_{n,{\bf k}}/\partial k_x$, $c_1(x)=x \ln{(x)}-(x+1) \ln{(x+1)}$ and $n_b(\epsilon)$ is the Bose-Einstein distribution. We define the magnon SNE angle as $\theta_{\rm SNE}=\sigma_{xy}/\sigma_{xx}$ proportional to $\alpha_0$.

\begin{figure}[t]
\includegraphics[width=8.6cm,height=8.6cm,keepaspectratio]{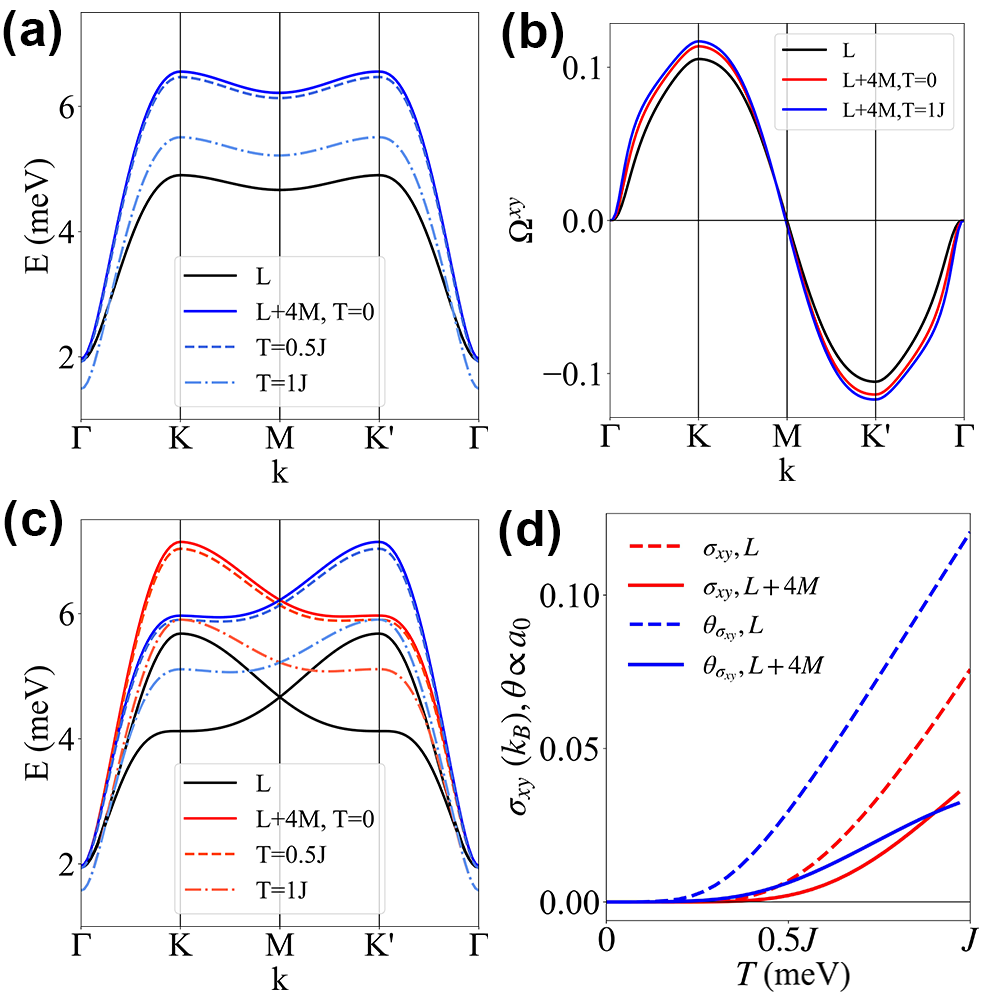}
\caption{(a,c) The magnon spectrum along the high symmetry path in the non-interacting LSWT (L) approximation (black lines) and in the presence of the four-magnon interactions (4M - colored lines), for (a) $D_z=0$ and (c) $D_z=0.1J$. (b) The Berry curvature along the high symmetry path for the case of non-interacting LSWT (black) and four-magnon interactions at $T=0$ (red) and at $T=1J$ (blue). (d) The spin Nernst conductivity and spin Nernst angle (arbitrary units) without interactions (dashed lines) and with four-magnon interactions (solid lines).}
\label{fig:fig2}
\end{figure}

The easy-axis anisotropy is set to $K'=0.2(2S-1)/S$, which stabilizes the magnon spectrum even in the presence of interactions, see Appendix D. In the calculations discussed below, we set $k_B=\hbar=1$ and vary the temperature $T$ and the values of the DMIs as a way to tune the interactions and the transport properties of the system. In Fig. \ref{fig:fig2}, we present the magnon spectrum in the non-interacting LSWT limit (black lines) in the absence (a) and presence (b) of an out-of-plane DMI, $D_z$. This DMI splits the degenerate bands at K and K' points, which causes an imbalance of magnon populations, thereby promoting a net magnon SNE \cite{cheng2016spin, zyuzin2016magnon,zhang2022perspective}.
Before addressing the impact of magnon interactions on the spectrum and transport properties, we stress that the perturbation theory discussed in the next section necessitates a stable magnetic order to start with. In our model, the mechanism that stabilizes the Neel antiferromagnetic order is the easy-axis anisotropy, while both the temperature and DMI destabilize it and can cause a phase transition either outside the antiferromagnetic phase or towards a noncollinear spin configuration. Without discussing the full phase diagram of the present honeycomb antiferromagnet, this stability condition constrains the temperature to remain well below the Neel temperature, i.e., typically $T<JS$. Similarly, DMI values of less than $D_{z,||}<0.2J$ should not cause a phase transition as long as a high easy-axis anisotropy is maintained, as discussed in Appendix D.

\section{Magnon Interactions\label{sectionIII}}

\begin{figure}[h]
\includegraphics[width=8.6cm,height=8.6cm,keepaspectratio]{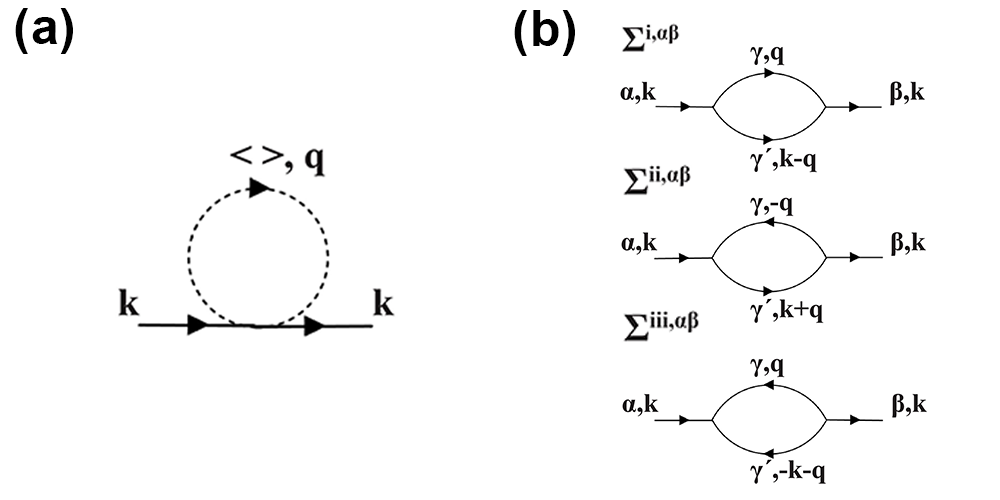}
\caption{(a) The Hartree diagram that describes the four-magnon interactions (b) The three Feynman diagrams that describe the different terms of self-energy in Eqs. \eqref{16.a}-\eqref{16.c}.}
\label{fig:fig1cd}
\end{figure}

So far, all calculations have been performed in the LSWT approximation, where the energy of the magnons is in the order of $O(S^1)$. Turning on magnon interactions, we consider the three-magnon interactions that are in the order of $O(S^{1/2})$ and the four-magnon interactions in the order $O(S^{0})$, where the appropriate diagrams are illustrated in Fig. \ref{fig:fig1cd}. To illustrate the impact of magnon interactions on the band structure and transport properties, we focus on the case of $S=3/2$. A discussion about this parameter and relevant materials realization is given in Section \ref{sectionIV}.

\subsection{Four-Magnon Interactions\label{sectionIIIA}}
The four-magnon interaction plays a similar role in the magnon gas as the Hartree-Fock correction in the electron gas. This interaction can be treated within a mean-field theory applied to the original $H^{(2)}$ Hamiltonian \cite{nolting2009quantum,rezende2019introduction}. Considering the second term in HP transformation of Eqs. \eqref{2.a}-\eqref{2.b}, we get the expanded four-magnon Hamiltonian is given by
\begin{widetext}
\begin{eqnarray}
H^{(4)}_{\rm HP}= -\frac{1}{4N}\sum_{{\bf k}_{1,2,3,4}}&&\bigg[J\left(f({\bf k}_4)a^{\dagger}_{{\bf k}_1}a_{{\bf k}_2}a_{{\bf k}_3}b_{{\bf k}_4}+f({\bf k}_{-4})a^{\dagger}_{{\bf k}_1}a^{\dagger}_{{\bf k}_2}a_{{\bf k}_3}b^{\dagger}_{{\bf k}_4}+f({\bf k}_{-2+3+4})a_{{\bf k}_1}b^{\dagger}_{{\bf k}_2}b_{{\bf k}_3}b_{{\bf k}_4}\right.\notag\\
&&\left.+f({\bf k}_{-2-3+4})a^{\dagger}_{{\bf k}_1}b^{\dagger}_{{\bf k}_2}b^{\dagger}_{{\bf k}_3}b_{{\bf k}_4}+4f({\bf k}_{4-3})a^{\dagger}_{{\bf k}_1}a_{{\bf k}_2}b^{\dagger}_{{\bf k}_3}b_{{\bf k}_4}\right)+K'\left(a^{\dagger}_{{\bf k}_1}a_{{\bf k}_2}a^{\dagger}_{{\bf k}_3}a_{{\bf k}_4}+(a\leftrightarrow b)\right) \notag\\ 
&&+D_z\bigg(g({\bf k}_{-4})a^{\dagger}_{{\bf k}_{1}}a_{{\bf k}_2}a_{{\bf k}_3}a^{\dagger}_{{\bf k}_4}+g({\bf k}_{-2+3+4})a^{\dagger}_{{\bf k}_1}a^{\dagger}_{{\bf k}_2}a_{{\bf k}_3}a_{{\bf k}_4}-(a\leftrightarrow b)\bigg)\bigg], \quad   \label{9}
\end{eqnarray}
\end{widetext}
where ${\bf k}_{1-2-3}={\bf k}_1-{\bf k}_2-{\bf k}_3$. In Appendix B, we describe how to derive the mean-field theory correction, which we add to the original LSWT. The final Hamiltonian reads
\begin{equation}
    H'=H^{(2)}+H^{(4)}(T), \label{10}
\end{equation}
\begin{equation}
    H^{(4)}(T)=  \sum_{\bf k} \Psi_{\bf k}^{\dagger} \begin{pmatrix}
    E^+({\bf k})&F^*({\bf k})\\
    F({\bf k})&E^-({\bf k})
    \end{pmatrix} \Psi_{\bf k},\label{11}
\end{equation}
which is quadratic in boson operators. The $H^{(4)}(T)$ term is the mean-field temperature correction to the energy and can be diagonalized in the same way as the LSWT Hamiltonian, $H^{(2)}$. This way, we obtain the renormalized energies and paraunitary matrix $\hat{T}_{\bf k}$ for the four-magnon interactions, which we can use to compute the spectrum, Berry curvature, and transport coefficients.

As presented in Appendix B, the term $H^{(4)}$ has two components, a thermal one, proportional to the magnon number that vanishes at $T=0$, and a quantum one, that is finite at $T=0$ \cite{nolting2009quantum,rezende2019introduction}. In ferromagnets, only the thermal component is present. The quantum contribution is a feature unique to antiferromagnets. Figure \ref{fig:fig2}(a) and (c) display the magnon spectrum at zero temperature (solid blue and red lines), in the absence and presence of out-of-plane DMI, respectively. One sees that the zero-point quantum fluctuations push the magnon spectrum upwards compared to the non-interacting case (solid black lines). In other words, at $T=0$, the four-magnon interactions increase the bandwidth and enhance the magnon velocity. As soon as $T>0$, the thermal contribution quicks in, reducing the effective magnetic exchange, thereby narrowing the magnon bandwidth, as exemplified for $T=0.5J$ (dashed lines) and  $T=1J$ (dotted dashed lines).

By comparing the Berry curvature along the high symmetry path in Fig. \ref{fig:fig2}(b), one sees that at its maximum the Berry curvature is enhanced by the quantum correction at $T=0$ (red line) in comparison with the non-interacting one (black line). Additionally, a small enhancement is obtained upon increasing the temperature up to $T=1J$ (blue line). However, these small increases in the Berry curvature cannot compensate for the much larger dampening of the energy spectrum. Consequently, the magnon transport is heavily damped by the four-magnon interactions, as shown in Fig. \ref{fig:fig2}(d). The interacting SNE conductivity (solid red line) and angle (solid blue line) are much smaller than the non-interacting ones (dashed lines). We conclude that not accounting for the four-magnon interactions, as usually achieved in the literature \cite{cheng2016spin,zyuzin2016magnon,zhang2022perspective}, leads to an overestimate of the magnon conductivity tensor at higher temperatures.

\subsection{Three-Magnon Interactions\label{sectionIIIB}}

We now turn our attention towards the three-magnon interactions that arise naturally as the coupling of out-of-plane and in-plane spin components $S^zS^{\pm}$ that the in-plane DMI introduces. These couplings are described by three-magnon operators and thus cannot be captured by LSWT. To properly account for these particle-non-conserving effects, we use quantum field theory techniques to compute the temperature-dependent self-energy of interacting magnons \cite{chernyshev2009spin,zhitomirsky2013colloquium}, and devise an effective Hamiltonian to describe the interacting magnon transport.

By applying the HP transformation to the in-plane DMI, we get the three-magnon term,
\begin{equation}
    H^{(3)}_{ij}= \sum_{\langle ij\rangle}\left( V_{ij}a_{i}b^{\dagger}_{j}b_{j} +V_{ij}a^{\dagger}_{i}a_{i}b^{\dagger}_{j}+h.c.\right), \label{12}
\end{equation}
where the real space term is given as $V_{ij}= D_{||}\sqrt{\frac{S}{2}}e^{i\phi_{ij}}$, where $\phi_{ij} = -\frac{2n\pi}{3}$ ($n$=1,2,0). We then perform the Fourier transform to study the magnon interaction in momentum-space,
\begin{equation}
    H^{(3)}_{\bf k}= \frac{1}{N^{1/2}}\sum_{\bf k,q}\left( V_{\bf k}a_{\bf k}b^{\dagger}_{\bf k+q}b_{\bf q} +V_{\bf k}a^{\dagger}_{\bf q}a_{\bf k+q}b^{\dagger}_{\bf k}+h.c.\right), \label{13}
\end{equation}
where $V_{\bf k}= D_{||}\sqrt{\frac{S}{2}}\sum_{\langle ij\rangle}e^{i(\phi_{ij}-{\bf k}\cdot{\bm\delta}^{(1)}_{ij})}$. We point out that these terms have the same geometrical form as the coupling proposed in Ref. \cite{matsumoto2020nonreciprocal} and are expected to cause a similar splitting in the magnon spectrum.

By using the Bogoliubov transformation described in Eq. \eqref{5}, the three-magnon Hamiltonian can be recast in the form \cite{chernyshev2009spin,zhitomirsky2013colloquium}
\begin{eqnarray}
H^{(3)}_{\bf k}&=& \frac{1}{N^{1/2}}\sum_{{\bf k,q}}\sum_{\lambda,\mu,\nu=\pm}\left(\Phi^{\lambda\mu\leftarrow \nu}_{{\bf k,q}\leftarrow {\bf k+q}}c_{\lambda,{\bf k}}^{\dagger}c_{\mu,{\bf q}}^{\dagger}c_{\nu,{\bf k+q}}
    \right.\nonumber\\&&\left.+\Phi^{0\leftarrow \lambda,\mu,\nu}_{{\bf 0}\leftarrow {\bf k,q,-(k+q)}}c_{\lambda,{\bf k}}c_{\mu,{\bf q}}c_{\nu,{\bf -(k+q)}}+h.c.\right), \label{14}
\end{eqnarray}
where the $\Phi$ matrices are the vertex terms and are given in detail in Appendix C. This is the general three-magnon term, where the first regular contribution can be found as well in ferromagnets, while the second contribution is unique to antiferromagnets due to the nature of the Bogoliubov transform. To determine the impact of this term, we need to find the interacting one-magnon Green's function. This is given by Matsubara Green's function,
\begin{eqnarray}
G_{{\bf k},\alpha\beta}(\tau)&\approx& G^{(0)}_{{\bf k},\alpha\beta}(\tau)-\frac{1}{2!} \int^{\beta}_0d\tau_1d\tau_2\label{15}\\
&&\times\big\langle\mathcal{T}\hat{H}^{(3)}(\tau_1)\hat{H}^{(3)}(\tau_2)c_{{\bf k},\alpha}(\tau)c^{\dagger}_{{\bf k},\beta}(0)\big\rangle, \notag
\end{eqnarray}
where the non-interacting Green's function is $G^{(0)}_{{\bf k},\alpha\beta}(\tau)=-\langle \mathcal{T}c_{{\bf k},\alpha}(\tau)c^{\dagger}_{{\bf k},\beta}(0) \rangle$, $\mathcal{T}$ is the time ordering and $\beta = 1/k_BT$. Higher orders in S and in $H_{\bf k}^{(3)}$ are not considered for simplicity.

We follow Refs. \cite{chernyshev2009spin,zhitomirsky2013colloquium,mook2021interaction,mahan2000many} to compute the temperature-dependent self-energies for the three-magnon Hamiltonian. The details are given in Appendix C. The expression for the self-energy of the interacting magnons has three contributions, corresponding to the regular (first two terms) and anomalous part (last term) of the Hamiltonian in Eq. \eqref{14},
\begin{widetext}
\begin{eqnarray}
\Sigma^{i,\alpha\beta}_{\bf k}(\epsilon,T)&=&\frac{1}{N}\sum_{\bf q}\sum_{\gamma,\gamma'=\pm} \frac{\Phi^{\alpha\leftarrow \gamma,\gamma'}_{{\bf k}\leftarrow {\bf q, k-q}}\Phi^{ \gamma,\gamma'\leftarrow \beta}_{{\bf q, k-q}\leftarrow {\bf k}}}{\epsilon+i\eta-\epsilon_{{\bf q},\gamma}-\epsilon_{{\bf k-q},\gamma'}}(n_b(\epsilon_{{\bf q},\gamma})+n_b(\epsilon_{{\bf k-q},\gamma'})+1),  \label{16.a}\\  
\Sigma^{ii,\alpha\beta}_{\bf k}(\epsilon,T)&=&\frac{1}{N}\sum_{\bf q}\sum_{\gamma,\gamma'=\pm}\frac{\Phi^{\gamma'\leftarrow \beta,\gamma}_{{\bf k+q}\leftarrow {\bf k,q}}\Phi^{ \alpha,\gamma\leftarrow \gamma'}_{{\bf k,q}\leftarrow {\bf k+q}}}{\epsilon+i\eta+\epsilon_{{\bf q},\gamma}-\epsilon_{{\bf k+q},\gamma'}} (n_b(\epsilon_{{\bf q},\gamma})-n_b(\epsilon_{{\bf k+q},\gamma'})), \label{16.b}\\
\Sigma^{iii,\alpha\beta}_{\bf k}(\epsilon,T)&=&-\frac{1}{N}\sum_{\bf q}\sum_{\gamma,\gamma'=\pm}\frac{\Phi^{0\leftarrow\alpha, \gamma,\gamma'}_{{\bf 0}\leftarrow {\bf k,q},-{\bf k-q}}\Phi^{\beta, \gamma,\gamma'\leftarrow 0}_{{\bf k,q,-k-q}\leftarrow {\bf 0}}}{\epsilon+\epsilon_{{\bf q},\gamma}+\epsilon_{{\bf -k-q},\gamma'}}(n_b(\epsilon_{{\bf q},\gamma})+n_b(\epsilon_{{\bf -k-q},\gamma'})+1). \label{16.c}
\end{eqnarray}
\end{widetext}
The terms $\Sigma^i_{\bf k}$ and $\Sigma^{ii}_{\bf k}$ are the same as the ones for ferromagnets, while the term $\Sigma^{iii}_{\bf k}$ is unique to antiferromagnets. The total self-energy is the sum $\hat{\Sigma}_{\bf k}(\epsilon,T)= \Sigma^i_{\bf k}(\epsilon,T) + \Sigma^{ii}_{\bf k}(\epsilon,T)+ \Sigma^{iii}_{\bf k}(\epsilon,T)$. By calculating the density of states of the processes, see Appendix D, we find that $\Sigma^i_{\bf k}$ dominates at high energies and low temperatures (in other words, it does not vanish at $T=0$), while $\Sigma^{ii}_{\bf k}$ dominates at low energies and high temperatures (it vanishes at $T=0$). The term $\Sigma^{iii}_{\bf k}$ plays a smaller role, especially at high temperatures.

\begin{figure}[t]
\includegraphics[width=8.6cm,height=8.6cm,keepaspectratio]{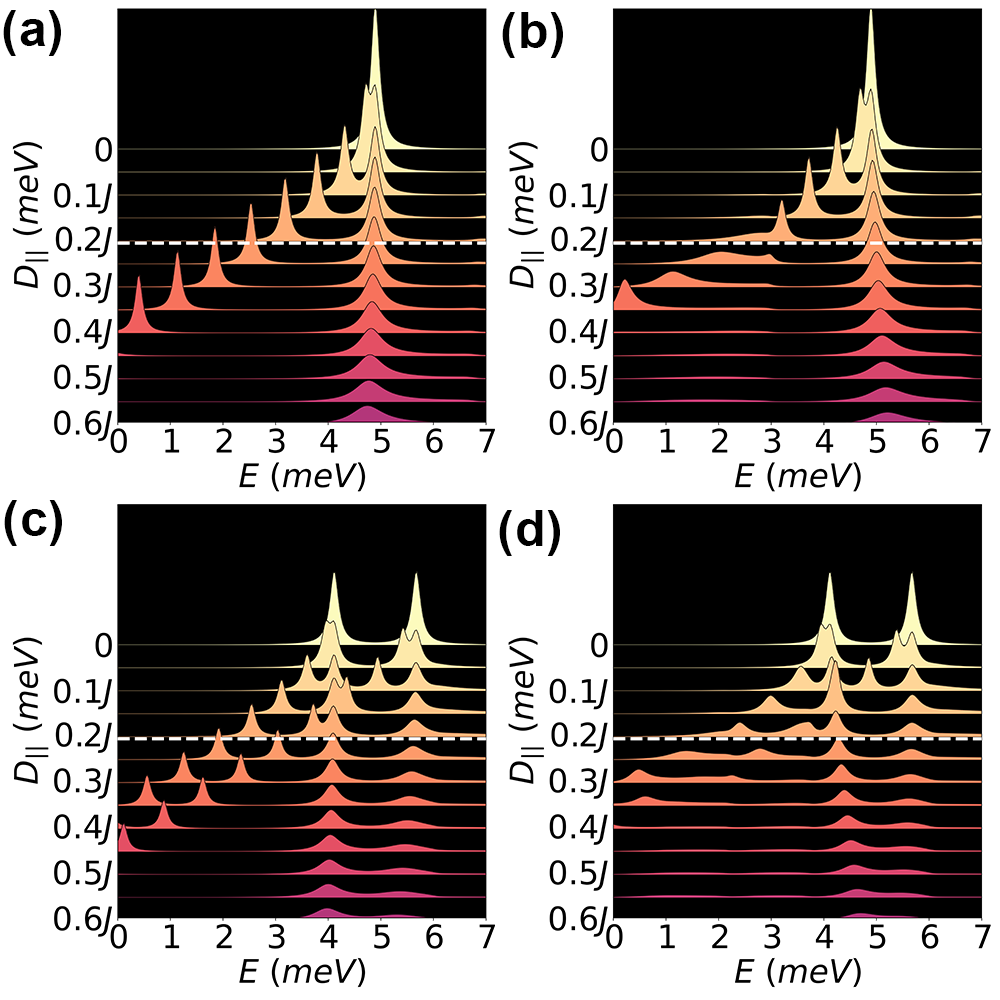}
\caption{The spectral function at K-point as a function of $D_{||}$. In (a) and (b) $D_z=0$, while for the lower panels, (c) and (d), $D_z=0.1J$. The left panels, (a) and (c), are at $T=0$ while the right panels, (b) and (d), are at $T=1J$. The white dashed line indicates $D_{||}=0.2J$ and the limit of well-defined magnonic excitations at any temperature. The numerical broadening is set to $\eta = 0.1$.}
\label{fig:fig3}
\end{figure}

The one-magnon spectral function is given by
\begin{equation}
    A_{\bf k}(\epsilon)=-\frac{1}{\pi}{\rm Im(Tr}[\hat{G}_{\bf k}(\epsilon)]), \label{18}
\end{equation}
where the interacting Green's function and the self-energy matrix have the form 
\begin{eqnarray}
\hat{G}_{\bf k}(\epsilon)&=&\left((\epsilon+i0^+)\hat{I}-\hat{\epsilon}_{\bf k}-\hat{\Sigma}_{\bf k}\right)^{-1},\\
    \hat{\Sigma}_{\bf k}(\epsilon,T) &=& \begin{pmatrix}
        \Sigma ^{++}&\Sigma ^{+-} \\
        \Sigma ^{-+}&\Sigma ^{--} \label{17}
    \end{pmatrix},
\end{eqnarray}
defined on the extended magnon basis. The spectral function at K-point, where the interactions are maximum, is reported in Fig. \ref{fig:fig3} for different values of the in-plane and out-of-plane DMI. A well-defined quasi-particle is characterized by a sharp Lorenztian peak, as we see for all subfigures in the absence of interactions ($D_{||} = 0$). As the interactions increase, the quasi-particle peak broadens and the magnonic excitations become ill-defined. This is due to the increased imaginary part of the self-energy that reduces the quasi-particle lifetime, while the real part causes the shift of the quasi-particle peak in the energy.

We can see the splitting of the degenerate bands, due to the real part of the self-energy, as the value of the in-plane DMI increases. This splitting is caused mainly by the off-diagonal terms of the self-energy $\Sigma^{\pm \mp}$ in the absence of $D_z$, as seen in Figs. \ref{fig:fig3}(a) and (b). We see that when the bands are already split, for $D_z\neq0$, a second splitting occurs leading to four bands, as shown in Figs. \ref{fig:fig3}(c) and (d), for $T=0$ and $T=1J$, respectively. The white lines indicate the value of $D_{||} = 0.2J$, which we consider the highest value of the three-magnon interaction parameter for which magnons remain well-defined quasi-particles at K-point (see also Appendix D).

\begin{figure}[t]
\includegraphics[width=8.6cm,height=8.6cm,keepaspectratio]{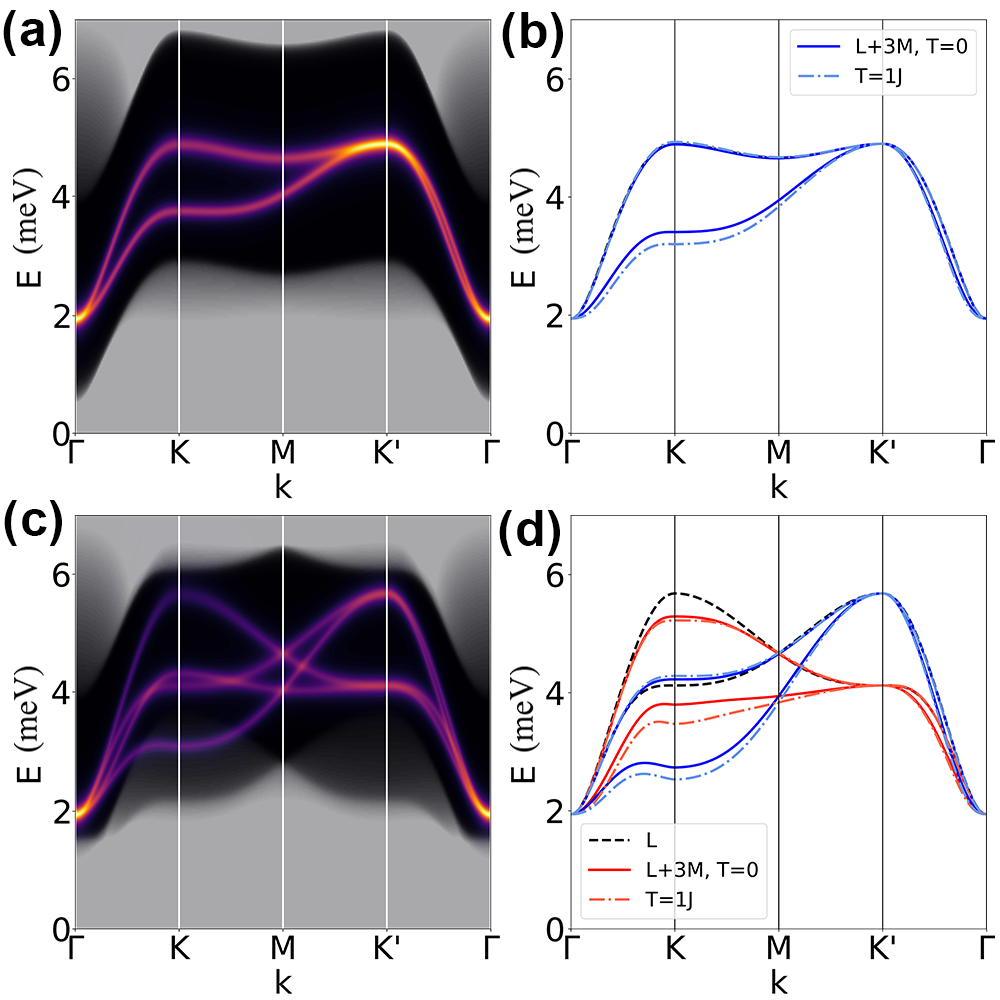}
\caption{Comparison between the spectral function $A_{\bf k}$ and the effective model (a),(b) for $D_z=0$ and (c),(d) for $D_z=0.1$. The spectral function is shown at $T=0$K, where the magnons are well-defined. In the effective model, the spectrum is shown for both $T=0$ (solid lines) and $T=1J$ (dashed-dotted lines). The opaque white overlay presents the two-magnon continuum of the models, see Appendix D.}
\label{fig:fig4}
\end{figure}

To calculate the magnon conductivities, we use an effective model that reproduces satisfactorily the spectral function. To do so, we follow the same procedure as in Section \ref{sectionII} [see Eq. \eqref{4}], and perform a second diagonalization of the interacting magnon Hamiltonian,
\begin{equation}
    \hat{F}_{\bf k}^\dagger(\hat{\epsilon}_{\bf k}+\tilde{\Sigma}_{\bf k}(\epsilon_{\bf k},T))\hat{F}_{\bf k}=\hat{\epsilon}_{\bf k}', \label{19}
\end{equation}
whereby $\tilde{\Sigma}_{\bf k}(\epsilon_{\bf k},T)$, here we assume only the Hermitian part of the self-energy, $\tilde{\Sigma}_{\bf k}=(\hat{\Sigma}_{\bf k}+\hat{\Sigma}_{\bf k}^{\dagger})/2$. In addition, we make the on-shell approximation, $\epsilon=\epsilon_{\bf k}$. To sum the energy matrix (diagonal) and self-energy matrix [block diagonal, where each block has the form of Eq. \eqref{17}], we need to use the extended basis. By following the diagonalization procedure described in Appendix A, we obtain both a numerical and an analytical solution to the problem. The comparison of the interacting spectral function (at $T=0$) is plotted against the interacting magnon spectrum at $T=0$ and $T=1J$ in Fig. \ref{fig:fig4}. In the absence of out-of-plane DMI, we find an excellent agreement between the spectral function and the effective model. In the presence of out-of-plane DMI, there are slight disagreements because the on-shell approximation does not hold as well as for the degenerate bands.

\begin{figure}[t]
\includegraphics[width=8.6cm,height=8.6cm,keepaspectratio]{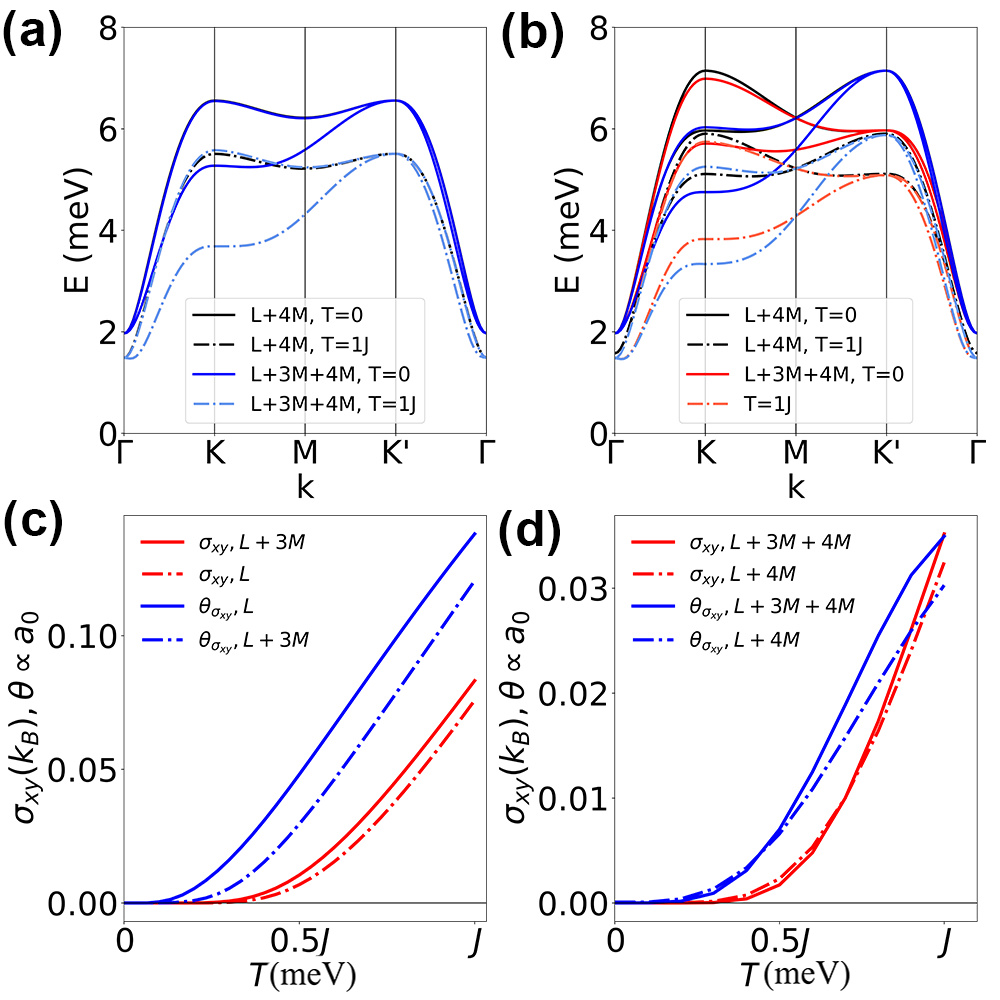}
\caption{\label{fig:epsart}(a,b) The magnon spectrum in the presence of both three- and four-magnon interactions as described by Eq. \eqref{20} (a) for $D_z=0$ and (b) $D_z=0.1J$. The black lines show the spectrum under only the four-magnon interactions, at $T=0$ (solid lines) and $T=1J$ (dashed lines). The colored lines show the spectrum for the full interactions, at $T=0$ (solid lines) and $T=1J$ (dashed lines). (c) The magnon SNE conductivity (red lines) and angle (blue lines) without interactions (dashed-dotted lines) and with three-magnon interactions (solid lines). (d) The magnon SNE conductivity (red lines) and angle (blue lines) for the four-magnon interactions (dashed-dotted lines) and full-magnon interactions (solid lines).}
\label{fig:fig5}
\end{figure}

By considering the total basis transformation $\hat{Q}_{\bf k}=\hat{F}_{\bf k}\hat{T}_{\bf k}$, we can calculate the total Berry curvature for the original and effective models from Eq. \eqref{6}. We numerically confirm that the Berry curvature is not affected by the three-magnon interactions and the Berry curvature of $\hat{Q}_{\bf k}$ states is the same as $\hat{T}_{\bf k}$ states. This is related to the fact that the three-magnon interactions lift the degeneracy of the magnon chirality $\pm$ but do not couple the ${\bf k}$ and $-{\bf k}$ sectors. Then, we calculate the longitudinal and transverse magnon spin transport in Fig. \ref{fig:fig5}(c) and we compare it with the non-interacting one. We numerically confirm that in the absence of out-of-plane DMI, the SNE is zero even in the presence of in-plane DMI, as the $\pm {\bf k}$ sectors are equal and $\sigma_{xy} = 0$. However, in the presence of both in-plane and out-of-plane DMI, both the SNE conductivity and angle are enhanced. The reason is that due to the symmetry of the in-plane DMI, more magnons are transported towards the transverse direction in comparison to the longitudinal one. We can see in Fig. \ref{fig:fig5}(c) that interactions enhance the SNE angle due to an increase of $\sigma_{xy}$ and a decrease of $\sigma_{xx}$.

\subsection{Full Magnon Interactions\label{sectionIIIC}}

In Section \ref{sectionIIIB}, we have seen that the temperature has a limited impact on the three-magnon interactions [see Figs. \ref{fig:fig4}(b,d)]. Now, let us consider the combined effect of three- and four-magnon interactions. This is achieved in two steps: (i) calculating the self-energy for the mean-field theory-corrected Hamiltonian, and then (ii) adding it to the corrected energies to find the effective model for the full interactions,
\begin{equation}
    \hat{F}^{\dagger'}_{\bf k}(\hat{\epsilon}_{\bf k}'+\tilde{\Sigma}_{\bf k}'(\epsilon_{\bf k}',T))\hat{F}_{\bf k}'=\hat{\epsilon}_{\bf k}''.\label{20}
\end{equation}
Here, the primed values are calculated for the Hamiltonian in Eq. \eqref{10}. In Fig. \ref{fig:fig5}(a) and (b), we present the magnon spectrum for the combination of three- and four-magnon interactions in the cases of $T=0$ and $T=1J$ (solid and dotted-dashed lines) in the (a) absence and (b) presence of $D_z$. The combination of the effects results in the combination of the non-reciprocal band splitting and large temperature-dependent renormalization, as discussed in details in the previous sections.

The next step is to calculate the magnon transport in the presence of both three- and four-magnon interactions. In Fig. \ref{fig:fig5}(d), we compute the transport for four-magnon (dotted dashed lines) and full-magnon interactions (solid lines). In Fig. \ref{fig:fig2}(d), we already observed that the four-magnon interactions tend to strongly suppress the SNE, whereas, on the contrary, Fig. \ref{fig:fig5}(c) indicates that the three-magnon interactions tend to enhance it. Nonetheless, the enhancement of the SNE induced by the three-magnon interaction does not overcome the dampening due to the four-magnon interactions, resulting in an overall reduction of the SNE signal by a factor of $\sim$3. In other words, although the prediction of the LSWT strongly overestimates the SNE, accounting for interactions does not suppress it.

\section{Material Consideration\label{sectionIV}}

The numerical calculations throughout this work have been performed for the case of $S=3/2$. Since the three- and four-magnon interactions discussed here are of the order of $O(S^{1/2})$ and $O(S^{0})$, respectively, they only dominate for small values of $S$; for $S>3/2$, the magnon interactions are suppressed and the LSWT becomes a reasonable approximation. That being said, the enhancement of the anomalous transport induced by the three-magnon interactions suggests that systems with both low spin and strong in-plane DMI could constitute promising platforms for the realization of interaction-enhanced SNE.

A first principle study \cite{bazazzadeh2021symmetry} on honeycomb antiferromagnets has identified the materials $\rm{MnPS_3, MnPSe_3}$ and $\rm{VPS_3}$ as potential Neel ordered antiferromagnets. While the Mn-based materials, with $S=5/2$, are already under investigation for their non-reciprocal band splitting \cite{wildes2021search} and magnon transport \cite{shiomi2017experimental,xing2019magnon}, the recently synthesized $\rm{VPS_3}$ \cite{liu2023probing} can present a viable alternative due to its smaller spin, $S=3/2$. Additionally, both the in-plane DMI and easy-axis anisotropy, which are crucial parameters for the onset of magnon interactions, can be manipulated via interfacial engineering \cite{hellman2017interface,doi:10.1021/acs.nanolett.2c00401}, a research direction that remained to be further developed.

\section{Discussion and conclusion\label{sectionV}}
 
Computing the magnon transport within the LSWT has led to the prediction of a wealth of intriguing transport effects and nontrivial topology in both ferromagnetic and antiferromagnetic systems \cite{mcclarty2022topological,bonbien2021topological,zhuo2023topological}. Nevertheless, although rather straightforward and well-established, this approach fails to account for several effects that are rooted in the nonlinear magnon interactions. This is particularly spectacular when considering the anomalous transport in ferromagnetic and antiferromagnetic lattices where, in the presence of inversion symmetry breaking, such interactions lead to substantial modification of the band structure and transport properties, including topological transitions \cite{mook2021interaction,lu2021topological,li2023temperature}. By focusing on the honeycomb antiferromagnet with in-plane DMI, we show that the three-magnon interactions can cause a non-reciprocal band splitting, even at $T=0$, and a small temperature renormalization at higher temperatures. Additionally, the four-magnon interactions cause a large renormalization of the magnon spectrum at higher temperatures. The combination of these two effects can cause an enhancement of transverse transport with expected advantages at low temperatures, while transport at higher temperatures may be rooted in magnon-phonon interactions, which we do not account for in this work.

The present work is limited to small values of the in-plane DMI, as it can be a source of a magnetic phase transition away from the collinear antiferromagnetic Neel state. Nonetheless, the nonreciprocity it induces is rather large, even at zero temperature, which is remarkable as it paves the way to non-reciprocal interacting transport in antiferromagnets. Additionally, since the in-plane DMI makes $S^z$ a non-conserved quantity, it breaks the hermicity of the magnon Hamiltonian, thereby shortening the lifetime of the magnonic quasi-particles, as indicated by the spectral function displayed in Fig. \ref{fig:fig3}. The non-hermitian contributions to the self-energy do not only limit the magnon lifetime though but can also influence the topology of the non-hermitian quantum states \cite{mcclarty2019non,deng2022non}. Modeling the transport properties resulting from such a "non-hermitian topology" requires the development of a quantum transport formalism that does not compute the response tensor from the effective quantum states, as performed in most magnonic studies including this one [see Eqs. \eqref{eq:7}-\eqref{eq:8}], but rather from the full interacting Green's functions of the system, as recently proposed \cite{Saleem2024Keldysh}.


While the in-plane DMI studied in the present work is one possible mechanism that introduces three-magnon interactions, noncollinear antiferromagnets are a natural platform for the onset of interactions. Such antiferromagnets have recently garnered attention in the context of antiferromagnetic spintronics \cite{jungwirth2016antiferromagnetic,baltz2018antiferromagnetic,jungwirth2018multiple,bonbien2021topological}. Predictions of unconventional magnon anomalous and spin Nernst effects \cite{mook2019thermal,li2020intrinsic,mook2019spin} and electron-magnon drag effects \cite{Cheng2020} have been recently made in a 120° antiferromagnetic kagome lattice in the context of LSWT. Because magnon interactions strongly influence the band structure and quasi-particle lifetime in this system \cite{chernyshev2009spin}, one cannot avoid accounting for these interactions to properly model these phenomena.
 
Finally, the present perturbation theory is based on the assumption that the magnons remain bosons up to relatively high temperatures and that HP expansion up to the fourth order in the magnon operators is sufficient. This question is central to the vast field of frustrated magnetism and we do not intend to provide an in-depth discussion of alternative theoretical approaches that could be used to address the problem of magnon interactions. Nonetheless, comparing such alternative formalisms with the perturbation theory considered in the present work, as proposed for instance in \cite{mook2021interaction}, would undoubtedly help better the condition of the emergence of these interactions and define upper and lower quantitative boundaries on their influence on magnon transport.


\begin{acknowledgments}
K. S. thanks Diego García Ovalle, Armando Pezo, and Luqman Saleem for fruitful discussions. K.S. and A.M. acknowledge support from the Excellence Initiative of Aix-Marseille Université–A*Midex, a French Investissements d’Avenir” program.
\end{acknowledgments}
\appendix

\section{Diagonalization of the Hamiltonian}
The diagonalization of an antiferromagnetic magnon Hamiltonian is a lot less straightforward than the case of ferromagnetic magnons due to its Bogoliubov-de-Gennes (BdG) bosonic nature. Throughout the main text, we use both a numerical (based on Ref. \cite{colpa1978diagonalization,toth2015linear}) and an analytical diagonalization (based on Ref. \cite{rezende2019introduction}). For the sake of completeness, we briefly discuss how the Hamiltonian \eqref{eq:1} is solved in the LSWT by using both methods.

The BdG Bosonic Hamiltonian has the general form
\begin{equation}
    H({\bf k}) = \begin{pmatrix}
    \hat{A}_{\bf k}&\hat{B}_{\bf k}\\
    \hat{B}_{\bf k}^{\dagger}&\hat{A}_{\bf -k}^T
    \end{pmatrix},
\end{equation}
and to diagonalize it numerically we can use the following scheme. We first Cholensky decompose the Hamiltonian as $H({\bf k})=\hat{K}^{\dagger}_{\bf k}\hat{K}_{\bf k}$. To perform the Cholensky decomposition, the Hamiltonian must be positive definite, which physically corresponds to the bosons having positive energies at all times. The energy matrix $\hat{E}_{\bf k}$ is obtained by finding the matrices $\hat{U}_{\bf k}$ that diagonalize $\hat{K}_{\bf k}\hat{G}_{\bf k}\hat{K}^{\dagger}_{\bf k}$  and satisfy the identity $\hat{U}_{\bf k}^{\dagger}\hat{K}_{\bf k}\hat{G}_{\bf k}\hat{K}_{\bf k}^{\dagger} \hat{U}_{\bf k} = \hat{L}_{\bf k}$ where $\hat{E}_{\bf k}=\hat{G}_{\bf k}\hat{L}_{\bf k}$. We can now calculate the matrices that diagonalize the original Hamiltonian as $\hat{T}_{\bf k}^{\dagger}\hat{H}_{\bf k}\hat{T}_{\bf k}= \hat{E}_{\bf k}$ and can be used to calculate the Berry curvature and other useful quantities from $\hat{T}_{\bf k}=\hat{K}^{-1}_{\bf k}\hat{U}_{\bf k}\sqrt{\hat{E}_{\bf k}}$.

This method can in theory numerically solve any antiferromagnetic magnonic problem, but an analytical solution is useful as discussed in the main text, especially for computing transport properties. We alternatively write the Hamiltonian of Eq. \eqref{3} with an extra element $C$ and express the diagonalization matrix in the extended basis $\Psi_{\bf k}=[a_{\bf k},b_{\bf k},a^{\dagger}_{\bf -k},b^{\dagger}_{\bf -k}]$,

\begin{eqnarray}
    H_{\bf k} &=& \begin{pmatrix}
    A_{\bf k}^+&C^*_{\bf k}&0&B^*_{\bf k}\\
    C_{\bf k}&A_{{\bf k}}^+&B^*_{\bf k}&0\\
    0&B_{\bf k}&A^-_{{\bf k}}&C_{\bf k}\\
    B_{\bf k}&0&C^*_{\bf k}&A^-_{{\bf k}}
    \end{pmatrix},\\
    \hat{T}_{\bf k} &=& \begin{pmatrix}
    T_{11}&T_{12}&T_{13}&T_{14}\\
    -T_{11}&T_{12}&-T_{13}&T_{14}\\
    T_{13}^*&T_{14}^*&T_{11}^*&T_{12}^*\\
    -T_{13}^*&T_{14}^*&-T_{11}^*&T_{12}^*
    \end{pmatrix},
\end{eqnarray}
where $A^\pm_{\bf k} = A_{\bf k}+ D_z g_{\pm {\bf k}}$. The $D_z $ term commutes with $\hat{T}_{\bf k}$ so it only affects the energies and not the $\hat{T}_{\bf k}$ itself, so we can suppress the $\pm$ index for calculating the analytical expression of $\hat{T}_{\bf k}$. The ${\bf k}$ subscripts of the elements of $\hat{T}_{\bf k}$ have been dropped for better readability.

We first present the solution for the simpler A-B model ($C_{\bf k}=0,\;A_{\bf k},B_{\bf k}\neq0$). The energies are given as $\epsilon_{\pm \bf k} = \omega_{\bf k} + D_z g_{\pm \bf k}$ where $\omega_{\bf k}=\sqrt{A_{\bf k}^2-|B_{\bf k}|^2}$ and $g_{\pm \bf k}=\pm g_{\bf k}$. The diagonalization matrix has the analytical solution
\begin{eqnarray}
&&T_{11}=\sqrt{\frac{A_{\bf k}+\omega_{\bf k}}{4\omega_{\bf k}}},\;T_{12}=-\sqrt{\frac{A_{\bf k}+\omega_{\bf k}}{4\omega_{\bf k}}},\nonumber\\
&&T_{13}=\sqrt{\frac{A_{\bf k}-\omega_{\bf k}}{4\omega_{\bf k}}}e^{i\phi_B},\;T_{14}=\sqrt{\frac{A_{\bf k}-\omega_{\bf k}}{4\omega_{\bf k}}}e^{i\phi_B},\nonumber
\end{eqnarray}
where $B_{\bf k}=|B_{\bf k}|e^{i\phi_B}$. This theory is given for the extended 4$\times$4 magnon basis. The same result can be given for the A-B model for the reduced 2$\times$2 basis as described in Section \ref{sectionII} of the main text and was used to derive the magnon SNE \cite{cheng2016spin,zyuzin2016magnon}.

We then give the solution for the A-C model ($B_{\bf k}=0,\;A_{\bf k},C_{\bf k}\neq0$) which is relevant to the effective model of our three-magnon interactions. The energies now read $\omega^{\pm}_{\bf k}=A_{\bf k}\pm |C_{\bf k}|$, and the elements of the diagonalization matrix $\hat{F}_{\bf k}$ read
\begin{eqnarray}
 F_{11}=\sqrt{\frac{A_{\bf k}-|C_{\bf k}|\chi_{\bf k}^-+\omega_{\bf k}^-}{4\omega_{\bf k}^-}},\;F_{13}=0,\nonumber\\F_{12}=\sqrt{\frac{A_{\bf k}+|C_{\bf k}|\chi_{\bf k}^++\omega_{\bf k}^+}{4\omega_{\bf k}^+}},\;F_{14}=0,\nonumber
\\ 
 \end{eqnarray}
where $\chi_{\bf k}^\pm= -1+2\left( \frac{\pm A_{\bf k}(2-\cos{\phi_C})+|C_{\bf k}|}{\pm A_{\bf k}+|C_{\bf k}|\cos{\phi_C}}\right)$, and $C_{\bf k}=|C_{\bf k}|e^{i\phi_C}$. For the analytical solution, we use the expression of the Berry curvature given in Eq. \eqref{6}. For the numerical solution, the alternative expression of Ref. \cite{mook2019thermal} should be used.

\section{Four-magnon mean-field theory}

We consider the expanded Fourier transformed Hamiltonian of Eq. \eqref{9} and perform a mean-field theory approximation. After performing a Bogoliubov transform, the expectation values of the pair correlators read \cite{rezende2019introduction,nolting2009quantum}
\begin{eqnarray}
    \langle a_{\bf q}^{\dagger}a_{\bf q} \rangle &=& |T_{11}|^2n(\epsilon_{+,\bf q})+|T_{14}|^2(n(\epsilon_{-,\bf q})+1),\\
    \langle b_{\bf q}^{\dagger}b_{\bf q} \rangle &=& |T_{12}|^2n(\epsilon_{-,\bf q})+|T_{13}|^2(n(\epsilon_{+,\bf q})+1),\\
    \langle a_{\bf q}b_{\bf q} \rangle &=& T_{11}T_{13}(n(\epsilon_{+,\bf q})+1) +T_{12}T_{14}n(\epsilon_{-,\bf q}),\\
    \langle a_{\bf q}^{\dagger}b^{\dagger}_{\bf q} \rangle &=& T_{11}^*T_{13}^*n(\epsilon_{+,\bf q}) +T_{12}^*T_{14}^*(n(\epsilon_{-,\bf q})+1).
\end{eqnarray}

We see that the mean-field terms have two parts, one that is proportional to the boson distribution and vanishes at $T=0$, and one that is finite even at $T=0$. The former is referred to as the thermal contribution, and the latter is the quantum fluctuation contribution. Using this mean-field expansion, the first term in Eq. \eqref{9} is written as a combination of two-magnon pair correlators,
\begin{eqnarray}
  f({\bf k}_4)a^{\dagger}_{{\bf k}_1}a_{{\bf k}_2}a_{{\bf k}_3}b_{{\bf k}_4}&=&2f({\bf k})\langle a^{\dagger}_{\bf q} a_{\bf q} \rangle a_{\bf k} b_{\bf k} \\
  &&+ 2f({\bf q})\langle a_{\bf q} b_{\bf q} \rangle a^{\dagger}_{\bf k} a_{\bf k}.\nonumber
\end{eqnarray}
This way, the Hamiltonian $H^{(4)}_{\rm HP}$ can be written in the form given by Eq. \eqref{11}, with 
%
\begin{widetext}
\begin{eqnarray}
E_{J}({\bf q},T) &=&-\frac{J}{2N}\sum_{\bf q} \left(f({\bf q}) \langle a_{\bf q}b_{\bf q} \rangle+f^*({\bf q})\langle a_{\bf q}^{\dagger}b^{\dagger}_{\bf q} \rangle+2f(0)\langle b_{\bf q}^{\dagger}b_{\bf q} \rangle \right),\\
E^{a}_{K}({\bf q},T) &=& -\frac{K}{N}\sum_{\bf q} \left( \langle a_{\bf q}^{\dagger}a_{\bf q} \rangle\right),\\
F_{J}({\bf k,q},T) &=&-\frac{J}{2N}\sum_{\bf q} \left(f({\bf k})(\langle a_{\bf q}^{\dagger}a_{\bf q} \rangle +\langle b_{\bf q}^{\dagger}b_{\bf q} \rangle )+2f({\bf q})\langle a_{\bf q}^{\dagger}b^{\dagger}_{\bf q} \rangle  \right),\\
H^{a}_{D_{z}}({\bf k,q},T) &=& -\frac{D_z}{2N}\sum_{\bf q} \left( (g({\bf k})+g({\bf q}))\langle a_{\bf q}^{\dagger}a_{\bf q} \rangle \right)
\end{eqnarray}
and $E^{\pm}_{\bf k}=E_{\bf k}\pm H_{\bf k}$. These terms are given for the $a$-sublattice, and we get the terms for $b$-sublattice by exchanging $a$ with $b$ operators.

\section{Three-magnon self-energies}

\subsection{Vertex terms}

By starting with Eq. \eqref{13}, where the magnons are written in the Fourier-transformed physical magnon basis, we perform the Bogoliubov transform that diagonalizes the magnons and transforms them in the eigenmagnon basis. We get two types of combinations of operators (and their hermitian conjugates): two magnons decaying into one, and three magnons decaying into the vacuum, which are proportional to vertex terms $\Phi$. The different vertex terms $\Phi$ are given here in detail:

\begin{eqnarray}
\Phi^{+\leftarrow ++}&=&\Phi^{+\leftarrow ++}_{-{\bf q\leftarrow k,-(k+q)}}+\Phi^{+\leftarrow ++}_{{\bf q\leftarrow -k,k+q}} =  V_{\bf k}T_{{\bf k},a+}T^*_{{\bf k+q},b+}T_{{\bf q},b+}+V_{\bf k}T^*_{{\bf q},a+}T_{{\bf k+q},a+}T^*_{{\bf k},b+}\\
\Phi^{-\leftarrow --} &=& \Phi^{-\leftarrow --}_{{\bf q\leftarrow -k,k+q}} + \Phi^{-\leftarrow --}_{{\bf -q\leftarrow k,-(k+q)}} =  V^*_{\bf k}T_{{\bf k},a-}^*T_{{\bf k+q},b-}T^*_{{\bf q},b-}+V^*_{\bf k}T_{{\bf q},a-}T^*_{{\bf k+q},a-}T_{{\bf k},b-}\\
\Phi^{+\leftarrow +-}&=& \Phi^{+\leftarrow +-}_{{\bf k\leftarrow -q, k+q}}+\Phi^{+\leftarrow +-}_{{\bf-(k+q)\leftarrow -q,- k} }+\Phi^{+\leftarrow +-}_{{\bf k+q\leftarrow k,q}}+\Phi^{+\leftarrow +-}_{{\bf-k\leftarrow q,-(k+q)}}\nonumber\\
&=&V^*_{\bf k}T_{{\bf k},a+}^*T_{{\bf k+q},b-}T_{{\bf q},b+}^*+V^*_{\bf k}T_{{\bf k},a-}^*T_{{\bf k+q},b+}T_{{\bf q},b+}^*+V^*_{\bf k}T_{{\bf q},a+}T_{{\bf k+q},a+}^*T_{{\bf k},b-}+V^*_{\bf k}T_{{\bf q},a+}T_{{\bf k+q},a-}^*T_{{\bf k},b+}\\
\Phi^{-\leftarrow +-}&=& \Phi^{-\leftarrow +-}_{{\bf k+q\leftarrow k,q} }+ \Phi^{-\leftarrow +-}_{{\bf -k\leftarrow -(k+q),q}}+    \Phi^{-\leftarrow +-}_{{\bf k\leftarrow k+q,-q }}+    \Phi^{-\leftarrow +-}_{{\bf -(k+q)\leftarrow -q,-k}}\nonumber\\
&=&V_{\bf k}T_{{\bf k},a+}T_{{\bf k+q},b-}^*T_{{\bf q},b-}+V_{\bf k}T_{{\bf k},a-}T_{{\bf k+q},b+}^*T_{{\bf q},b-}+V_{\bf k}T_{{\bf q},a-}^*T_{{\bf k+q},a+}T_{{\bf q},a-}^*+V_{\bf k}T_{{\bf q},a-}^*T_{{\bf k+q},a-}T_{{\bf k},b+}^*\\
\Phi^{0\leftarrow ++-}&=&\Phi^{0\leftarrow ++-}_{k,-(k+q),q}+    \Phi^{0\leftarrow ++-}_{-k,(k+q),-q}=
V_{\bf k}T_{{\bf k},a+}T_{{\bf k+q},b+}^*T_{{\bf q},b-}+V_{\bf k}T_{{\bf q},a-}^*T_{{\bf k+q},a+}T_{{\bf k},b+}^*\\
\Phi^{0\leftarrow +--}&=&    \Phi^{0\leftarrow +--}_{{\bf -q, k+q,- k}}+    \Phi^{0\leftarrow +--}_{{\bf q,-(k+q),k}}= V^*_{\bf k}T_{a-}^*T_{{\bf k+q},b-}T_{{\bf q},b+}^*+V^*_{\bf k}T_{{\bf q},a+}T_{{\bf k+q},a-}^*T_{{\bf k},b-}.
\end{eqnarray}
The summation over $\pm$ in the main text is taken only over the combination given in $\Phi$ as some combinations are not permitted by the nature of the Bogoliubov transform.

\subsection{Self energies}

We now calculate the self-energies for the general Hamiltonian Eq. \eqref{14}. In the limit of $T=0$, they were given in Ref. \cite{chernyshev2009spin, zhitomirsky2013colloquium}. To obtain the self-energies in Eqs. \eqref{16.a}-\eqref{16.c}, we inject Eq. \eqref{14} in Eq. \eqref{15}, i.e.,
\begin{equation}
    \big\langle\mathcal{T}\hat{H}^{(3)}(\tau_1)\hat{H}^{(3)}(\tau_2)c_{{\bf k},\alpha}(\tau)c^{\dagger}_{{\bf k},\beta}(0)\big\rangle,
\end{equation}
which can be separated into two different terms,

\begin{eqnarray}
&&\Phi^{\lambda_1 \mu_1 \leftarrow \nu_1}_{{\bf p}_1\leftarrow {\bf k}_1, {\bf q}_1 }  \Phi^{\nu_2 \leftarrow \lambda_2\mu_2 }_{ {\bf p}_2 \leftarrow {\bf k}_2, {\bf q}_2}  \big\langle\mathcal{T}c^{\dagger}_{{\bf k}_1 ,\lambda_1}(\tau_1)c^{\dagger}_{{\bf q}_1 ,\mu_1}(\tau_1)c_{{\bf p}_1,\nu_1}(\tau_1)c_{{\bf k}_2 ,\lambda_2}(\tau_2)c_{{\bf q}_2 ,\mu_2}(\tau_2)c^{\dagger}_{{\bf p}_2,\nu_2}(\tau_2)c_{{\bf k},\alpha}(\tau)c^{\dagger}_{{\bf k},\beta}(0)\big\rangle,\label{C8}\\&&
\Phi^{0\leftarrow \lambda_1 \mu_1 \nu_1}_{0\leftarrow {\bf k}_1, {\bf q}_1 ,-{\bf p}_1} \Phi^{\lambda_2\mu_2 \nu_2 \leftarrow 0}_{ {\bf k}_2, {\bf q}_2, -{\bf p}_2 \leftarrow 0} \big\langle\mathcal{T}c_{{\bf k}_1 ,\lambda_1}(\tau_1)c_{{\bf q}_1,\mu_1 }(\tau_1)c_{-{\bf p}_1,\nu_1}(\tau_1)
c^{\dagger}_{{\bf k}_2 ,\lambda_2}(\tau_2)c^{\dagger}_{{\bf q}_2,\mu_2}(\tau_2)c^{\dagger}_{-{\bf p}_2,\nu_2}(\tau_2)c_{{\bf k},\alpha}(\tau)c^{\dagger}_{{\bf k},\beta}(0)\big\rangle.\label{C9}
\end{eqnarray}

Using Wick's theorem, we get three forms of diagrams: the disconnected ones which are eliminated, the tadpole diagrams that integrate to zero, and the bubble diagrams which contribute to our three-magnon interactions, as illustrated in Fig. \ref{fig:fig1}(c). We express these terms as a function of the non-interacting Green's functions,
\begin{eqnarray}
   && \Phi^{\alpha\mu\leftarrow \nu}_{{\bf k,q\leftarrow k+q}} \Phi^{\nu\leftarrow\beta\mu }_{{\bf k+q\leftarrow k,q}} G_{\alpha}({\bf k},\tau-\tau_1)G_{\lambda}({\bf k},\tau_2-\tau_1)G_{\nu}({\bf k+q},\tau_1-\tau_2)G_{\beta}({\bf k},\tau_2),\\&&
    \Phi^{\lambda\alpha\leftarrow \nu}_{{\bf k,q\leftarrow k+q}} \Phi^{\nu\leftarrow\lambda\beta }_{{\bf k+q\leftarrow k,q}} G_{\alpha}({\bf k},\tau-\tau_1)G_{\mu}({\bf q},\tau_2-\tau_1)G_{\nu}({\bf k+q},\tau_1-\tau_2)G_{\beta}({\bf k},\tau_2),\\&&
    \Phi^{\lambda\mu\leftarrow \alpha}_{{\bf k,q\leftarrow k+q}} \Phi^{\beta\leftarrow\lambda \mu}_{{\bf k+q\leftarrow k,q}} G_{\alpha}({\bf k+q},\tau-\tau_2)G_{\mu}({\bf q},\tau_2-\tau_1)G_{\lambda}({\bf k},\tau_2-\tau_1)G_{\beta}({\bf k+q},\tau_1),
\end{eqnarray}
and for Eq. \eqref{C9},
\begin{eqnarray}
    &&\Phi^{\alpha\mu\nu\leftarrow 0}_{{\bf k,q, -(k+q)\leftarrow 0}} \Phi^{0\leftarrow\beta\mu\nu }_{{\bf 0\leftarrow k,q, -(k+q)}} G_{\alpha}({\bf k},\tau-\tau_2)G_{\lambda}({\bf q},\tau_1-\tau_2)G_{\nu}({\bf -(k+q)},\tau_1-\tau_2)G_{\beta}({\bf k},\tau_2),\\&&
    \Phi^{\lambda\alpha\nu\leftarrow 0}_{{\bf k,q, -(k+q)\leftarrow 0}} \Phi^{0\leftarrow\lambda\beta\nu }_{{\bf 0\leftarrow k,q, -(k+q)}} G_{\alpha}({\bf k},\tau-\tau_2)G_{\mu}({\bf q},\tau_1-\tau_2)G_{\nu}({\bf -(k+q)},\tau_1-\tau_2)G_{\beta}({\bf k},\tau_2),\\&&
    \Phi^{\lambda\mu\alpha\leftarrow 0}_{{\bf k,q, -(k+q)\leftarrow 0}} \Phi^{0\leftarrow\lambda\mu\beta }_{{\bf 0\leftarrow k,q, -(k+q)}} G_{\alpha}({\bf -(k+q)},\tau-\tau_2)G_{\lambda}({\bf q},\tau_1-\tau_2)G_{\mu}({\bf k},\tau_1-\tau_2)G_{\beta}({\bf -(k+q)},\tau_2).
\end{eqnarray}

Then, using the Fourier transform of the Green's function 
$G^{(0)}_{{\bf k},\alpha\beta}(\tau_1-\tau_2)=\frac{1}{\beta}\sum_{\omega_n}e^{-i\omega_n(\tau_1-\tau_2)}G^{(0)}_{{\bf k},\alpha\beta}(\omega_n)$, we obtains terms of the following forms for Eq. \eqref{C8}

\begin{eqnarray}
&&\Phi^{\alpha\mu\leftarrow \nu}_{{\bf k,q\leftarrow k+q}} \Phi^{\nu\leftarrow\beta\mu }_{{\bf k+q\leftarrow k,q}} G_{\alpha}({\bf k},k_n)G_{\lambda}({\bf q}, q_n)G_{\nu}({\bf k+q},k_n+q_n)G_{\beta}({\bf k},k_n),\\
    &&\Phi^{\lambda\alpha\leftarrow \nu}_{{\bf k,q\leftarrow k+q}} \Phi^{\nu\leftarrow\lambda\beta }_{{\bf k+q\leftarrow k,q}} G_{\alpha}({\bf k},k_n)G_{\mu}({\bf q},q_n)G_{\nu}({\bf k+q},k_n+q_n)G_{\beta}({\bf k},k_n),\\
    &&\Phi^{\lambda\mu\leftarrow \nu}_{{\bf k,q\leftarrow k+q}} \Phi^{\beta\leftarrow\lambda \mu}_{{\bf k+q\leftarrow k,q}} G_{\alpha}({\bf k+q},k_n)G_{\mu}({\bf q},q_n)G_{\lambda}({\bf k},k_n-q_n)G_{\beta}({\bf k+q},k_n),\\
&&\Phi^{\alpha\mu\nu\leftarrow 0}_{{\bf k,q, -(k+q)\leftarrow 0}} \Phi^{0\leftarrow\beta\mu\nu }_{{\bf 0\leftarrow k,q, -(k+q)}} G_{\alpha}({\bf k},k_n)G_{\lambda}({\bf q},q_n)G_{\nu}({\bf -(k+q)},-(k_n+q_n))G_{\beta}({\bf k},k_n),\\
 &&\Phi^{\lambda\alpha\nu\leftarrow 0}_{{\bf k,q, -(k+q)\leftarrow 0}}\Phi^{0\leftarrow\lambda\beta\nu }_{{\bf 0\leftarrow k,q, -(k+q)}} G_{\alpha}({\bf k},k_n)G_{\mu}({\bf q},q_n)G_{\nu}({\bf -(k+q)},-(k_n+q_n))G_{\beta}({\bf k},k_n),\\
  &&\Phi^{\lambda\mu\alpha\leftarrow 0}_{{\bf k,q, -(k+q)\leftarrow 0}} \Phi^{0\leftarrow\lambda\mu\beta }_{{\bf 0\leftarrow k,q, -(k+q)}} G_{\alpha}({\bf -(k+q)},k_n)G_{\lambda}({\bf q},q_n)G_{\mu}({\bf k},-(k_n+q_n))G_{\beta}({\bf -(k+q)},k_n).\end{eqnarray}
\end{widetext}
Then we perform a Matsubara zero-frequency summation to get the self-energies of Eqs. \eqref{16.a}-\eqref{16.c} \cite{mahan2000many}. The first two equations sum up to the $\Sigma^{ii}_{\bf k}$ term, and the third to the $\Sigma^i_{\bf k}$ term. The last three equations give the same self-energy contribution $\Sigma^{iii}_{\bf k}$. $\Sigma^i_{\bf k}$ and $\Sigma^{ii}_{\bf k}$ are also present in the case of the ferromagnetic magnons while $\Sigma^{ii}_{\bf k}$ is unique to antiferromagnetic magnons.

\section{Easy Axis Stabilization}
\twocolumngrid\bigskip

\begin{figure}[htp]
\includegraphics[width=8.6cm,height=8.6cm,keepaspectratio]{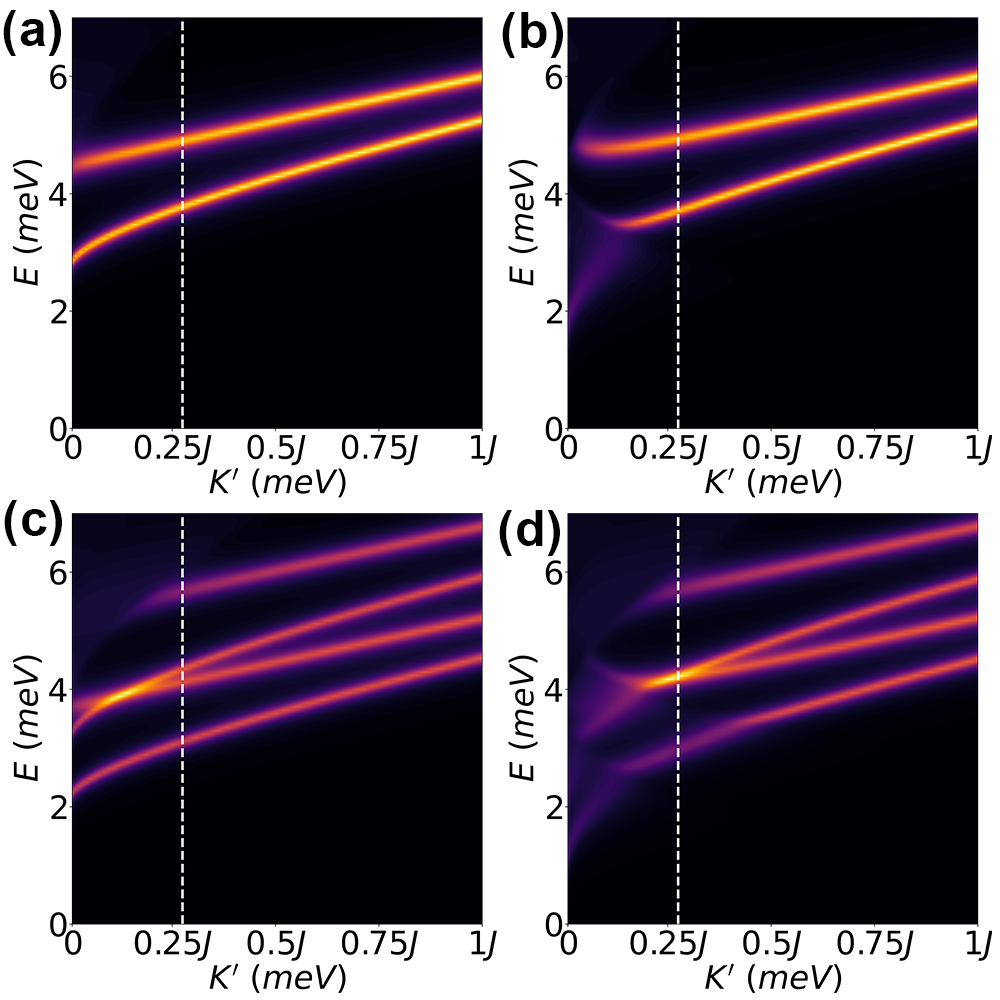}
\caption{Spectral function at K-point as a function of $K'$, for (a,b) $D_z=0$ and (c,d) $D_z=0.1J$, and for (a,c) $T=0$ and (b,d) $T=1J$. The white dashed line indicates $K'=0.2$ which corresponds to the limit of well-defined magnon excitations at any temperature. }
\label{fig:fig6}
\end{figure}
\begin{figure}[t]
\includegraphics[width=8.6cm,height=8.6cm,keepaspectratio]{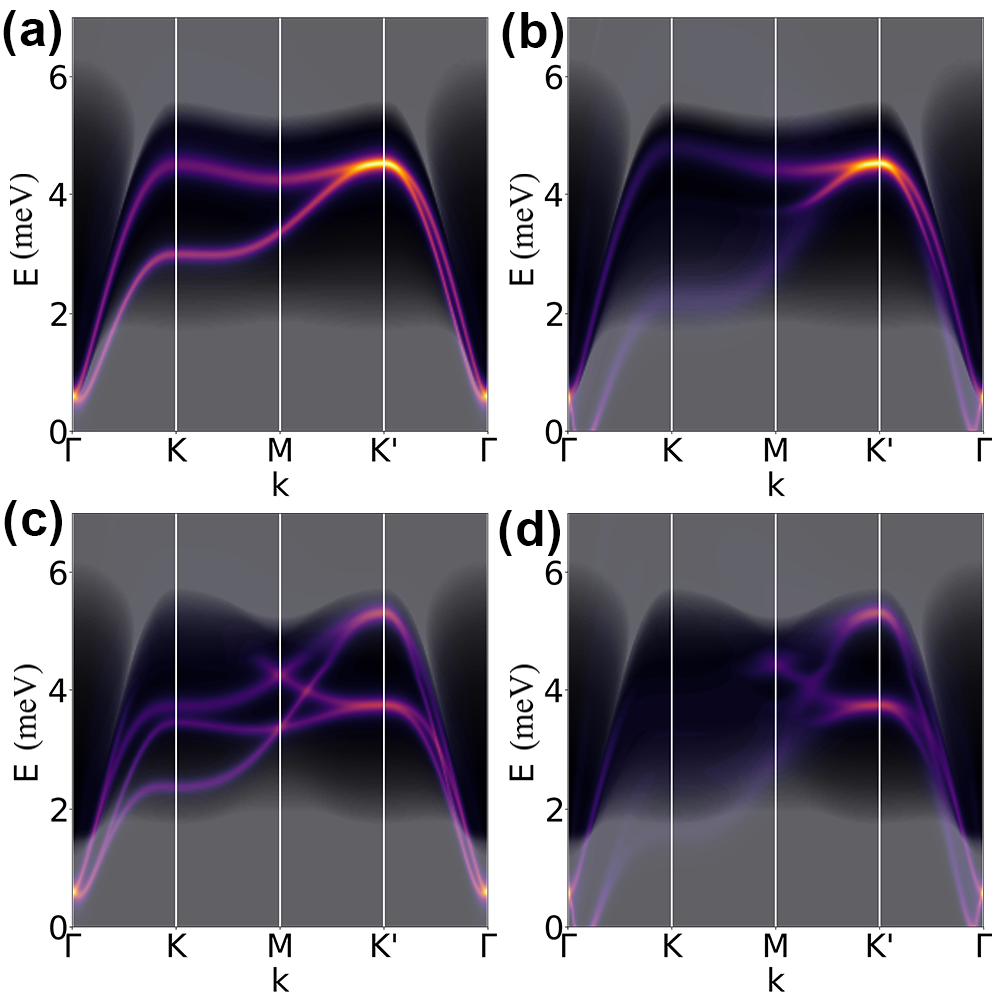}
\caption{Spectral function along the high-symmetry path of the Brillouin zone for $K'=0.02(2S-1)/S$, for (a,b) $D_z=0$ and (c,d) $D_z=0.1J$, and for (a,c) $T=0$ and (b,d) $T=1J$. The opaque white overlay presents the continuum of the models.}
\label{fig:fig7}
\end{figure}

It is well known that in magnon-magnon interactions the continuum plays an important role in the "good" definition of magnons. In other words, if the magnon continuum is too close to the (interaction-renormalized) one-magnon bands, these magnons are ill-defined. In ferromagnets, field-polarization by a magnetic field was considered to push the continuum far from the single-particle spectrum \cite{mook2021interaction}. In antiferromagnets, the magnetic field can cause a phase transition to spin-flop phases and ferromagnetism and thus destabilize the antiferromagnetic order. Instead, we consider the easy-axis anisotropy as the stabilization mechanism that can be tuned along with the in-plane DMI by interfacial engineering, as mentioned in Section \ref{sectionIV}.

In Fig. \ref{fig:fig6}, we compute the spectral function at the K-point as a function of the easy-axis anisotropy. We see that for small anisotropy, the magnons are not well-defined, resulting in a blurry signature. The white dashed line indicates the value $K' = 0.2J$, which we adopt throughout most of the work, as the lower limit of the easy-axis anisotropy where the magnons are well-defined.

The magnon interactions are kinematically allowed only in certain energies and momenta combinations \cite{chernyshev2009spin,zhitomirsky2013colloquium}. A good way to estimate whether such interactions are allowed is the two-magnon Density of States (DOS) \cite{mook2021interaction},
\begin{equation}
    D^i(\epsilon)_{\bf k}=\sum_{\gamma\gamma'}\sum_{\bf q}\delta (\epsilon-\epsilon_{{\bf q},\gamma}-\epsilon_{{\bf k- q},\gamma'}).
\end{equation}
\begin{equation}
    D^{ii}(\epsilon)_{\bf k}=\sum_{\gamma\gamma'}\sum_{\bf q}\delta (\epsilon+\epsilon_{{\bf q},\gamma}-\epsilon_{{\bf k+q},\gamma'}).
\end{equation}
\begin{equation}
    D^{iii}(\epsilon)_{\bf k}=\sum_{\gamma\gamma'}\sum_{\bf q}\delta (\epsilon+\epsilon_{{\bf q},\gamma}+\epsilon_{{\bf -k- q},\gamma'}).
\end{equation}
where $\delta(x)$ is the Dirac delta function, the indices $i-iii$ correspond to the denominators in Eqs. \eqref{16.a}-\eqref{16.c} and the total DOS is $D_{\bf k}(\epsilon)=D_{\bf k}^{i}(\epsilon)+D_{\bf k}^{ii}(\epsilon)+D_{\bf k}^{iii}(\epsilon)$.

The opaque white overlay in Figs. \ref{fig:fig4} and \ref{fig:fig7} represent the two-magnon DOS. When the spectral functions cross with the DOS, magnon decay occurs and the magnon quasiparticle lines are blurred. This is better illustrated in Fig. \ref{fig:fig7} where a smaller easy-axis anisotropy is chosen for the same values of DMI as Fig. \ref{fig:fig5}. It is obvious that in this case, the three-magnon interactions make the magnon quasiparticles ill-defined. For this reason, a high easy-axis anisotropy is necessary to stabilize the antiferromagnetic magnons under strong interactions. We comment that if the magnons are kinematically protected against three-magnon interactions, they should also be protected against four- and higher-order magnon interactions \cite{zhitomirsky2013colloquium,harris1971dynamics}.

\bibliography{bibliography}

\end{document}